\pdfoutput=1
\documentclass[conference]{IEEEtran}

\usepackage{amsmath,amssymb,amsthm}
\usepackage{booktabs}
\usepackage{cite}
\usepackage{enumitem}
\usepackage{graphicx}
\usepackage{array}
\usepackage{multirow}
\usepackage{placeins}
\usepackage{url}

\setlist[itemize]{leftmargin=*,nosep}
\setlist[enumerate]{leftmargin=*,nosep}

\newtheorem{theorem}{Theorem}
\newcounter{algorithm}

\makeatletter
\newcommand{\inlinefigcaption}[1]{%
  \@makecaption{\figurename~\thefigure}{#1}%
}
\makeatother

\newcommand{\method}{LD-Leiden}
\newcommand{\graph}{G}
\newcommand{\batch}{\Delta}
\newcommand{\partn}{\mathcal{P}}
\newcommand{\quality}{Q}
\newcommand{\frontier}{\mathcal{A}}
\newcommand{\hier}{\mathcal{H}}

\newcommand{\algbracket}[1]{\ensuremath{\vcenter{\offinterlineskip
  \hbox to .42em{\rule{.42em}{.24pt}\hfil}
  \hbox to .42em{\rule{.24pt}{#1}\hfil}
  \hbox to .42em{\rule{.42em}{.24pt}\hfil}}}}
\newcommand{\algbrace}[2]{\multirow{#1}{*}{\smash{\algbracket{#2}}}}
\newcommand{\algblock}[2]{\multirow{#1}{*}{\rotatebox[origin=c]{90}{\tiny\textsc{#2}}}}

\title{\method: Local Parallel Community Detection in Large Dynamic Networks}

\author{
\IEEEauthorblockN{Grigoriy Bokov\thanks{Corresponding author: Grigoriy Bokov (bokovgrigoriy@gmail.com).}}
\IEEEauthorblockA{Lomonosov \\
Moscow State University\\
Moscow, Russia\\
1 Leninskiye Gory, 119991\\
bokovgrigoriy@gmail.com}
\IEEEauthorblockN{Stanislav Moiseev}
\IEEEauthorblockA{Lomonosov \\
Moscow State University\\
Moscow, Russia\\
1 Leninskiye Gory, 119991\\
stanislav.moiseev@gmail.com}
\and
\IEEEauthorblockN{Aleksandr Konovalov}
\IEEEauthorblockA{Lomonosov \\
Moscow State University\\
Moscow, Russia\\
1 Leninskiye Gory, 119991\\
alexandr.konoval@gmail.com}
\IEEEauthorblockN{Ivan Safonov}
\IEEEauthorblockA{National Research University\\
Higher School of Economics\\
Moscow, Russia\\
11 Pokrovksy Bulvar, 109028\\
isafonov27@gmail.com}
\and
\IEEEauthorblockN{Anna Uporova}
\IEEEauthorblockA{Lomonosov \\
Moscow State University\\
Moscow, Russia\\
1 Leninskiye Gory, 119991\\
annuporova2003@gmail.com}
\IEEEauthorblockN{Alexander Radionov}
\IEEEauthorblockA{Moscow Infocommunication \\
Technology Laboratory\\
Moscow, Russia\\
1-75b Leninskiye Gory, 119234\\
alex.radionov89@gmail.com}
}

\IEEEoverridecommandlockouts

\begin{document}
\maketitle
\raggedbottom

\begin{abstract}
Dynamic community detection must update high-quality modularity partitions
after edge batches, yet full Leiden reruns make small changes scale with the
whole snapshot. Existing dynamic methods reduce work but often alter Leiden
refinement, keep limited hierarchy state, or restrict graph support. This paper
presents \method, a local dynamic Leiden method for weighted directed and
undirected graphs that preserves the move-refine-aggregate pipeline and updates
only repaired affected regions. Its novelty is the combination of an
affected-frontier rule after statistic repair, exact subtract-add aggregate
repair, and conflict-filtered parallel local moves; together these mechanisms
bound update cost by the visited frontier rather than the full graph. On real
streams and streamed static graphs with up to 214M vertices and 3.30B edges,
\method{} is 48.77$\times$ faster than warm-started Leidenalg in 100-batch runs
while preserving a 0.996 final modularity ratio. On the common undirected
benchmark set, it is 6.94$\times$ faster than DF-Leiden and 9.73$\times$ faster
than NetworKit while obtaining higher final modularity; synthetic sequences
support the predicted local edge-volume scaling.
\end{abstract}


\section{Introduction}

Large networks are rarely static. Social interactions, citation links,
transaction graphs, road updates, and communication networks evolve through
small batches of edge insertions, deletions, and weight changes. Community
detection is a standard tool for summarizing and exploring network structure
~\cite{Newman2003,Porter2009,Fortunato2010}. Modularity optimization provides
a common objective for this task through its null-model interpretation, standard
implementations, and established large-scale benchmarks
~\cite{NewmanGirvan2004,Brandes2008}. Louvain made greedy modularity
practical on large graphs~\cite{Blondel2008}; Leiden improved the optimization
path by refining communities before aggregation, thereby avoiding poorly
connected communities~\cite{Traag2019}.

The dynamic version of this problem adds a state-maintenance requirement. A
full Leiden rerun after every update provides a conservative quality
reference, but it repeatedly scans large parts of the graph even when the
batch touches only a small region.
Existing dynamic methods reuse the previous partition, screen unaffected
vertices, or restrict updates to a dynamic frontier~\cite{Rossetti2018,
Zhuang2019,Zarayeneh2021,Sahu2024Louvain,Sahu2024Leiden}. These mechanisms
reduce repeated work, yet the methods that can be run under a common protocol either
remain Louvain-like, drop support for directed graphs, or do not maintain a
Leiden refinement hierarchy incrementally. The resulting algorithmic problem
is to preserve the refinement and aggregation state used by Leiden while
avoiding a full-snapshot scan after every batch.

To address this problem, \method{} is designed as a local parallel Leiden
algorithm for dynamic weighted graphs. \method{} keeps the Leiden move-refine-aggregate
structure and restricts each stage to the region whose modularity context may
have changed. Its first contribution is an affected-frontier update rule tied
to an explicit one-step modularity-gain formula, so candidate moves are
evaluated from repaired current-state weights, community totals, and the
global edge-weight total. Empirically, reducing the batch size by a factor of
ten improves \method{} runtime
by 4.91$\times$ on average, far more than for the full-snapshot baselines.

The second contribution is an incremental hierarchy for dynamic Leiden.
Instead of rebuilding aggregate graphs after every batch, \method{} stores
ground nodes, refined nodes, parent links, community totals, and aggregate
edge weights at each hierarchy level. A batch repairs only the hierarchy paths
containing affected communities; the local subtract-add repair gives the same
aggregate weights as a full rebuild on all changed supernodes.

The third contribution is a parallel local-move scheme. Threads search for
positive-gain candidates independently, and a greedy decoupling filter accepts
a mini-batch in which no community is both an emitter and an acceptor.
The filter gives a deterministic consistency rule for additive community
updates; the measured shared-memory study gives a 5.22$\times$ mean speedup
from 1 to 64 threads.

Finally, the paper evaluates the method on real and synthetic streams. In
100-batch real-graph runs, \method{} is 48.77$\times$ faster than
warm-started Leidenalg while preserving a 0.996 final modularity ratio. On the
common undirected subset, it is 43.45$\times$, 9.73$\times$, and
6.94$\times$ faster than Grappolo, NetworKit, and DF-Leiden, respectively,
while obtaining higher final modularity and using 50.9\%--64.3\% of baseline
resident memory on matched graphs. Its final modularity varies by only 0.38\%
on average across 1--64 thread runs. On synthetic SBM-style dynamic graph
sequences, \method{} retains planted-community recovery in identifiable
regimes while the update time grows with local edge volume. The
evaluation also states the operating limits: real-graph quality is assessed by
modularity rather than partition identity, all experiments use $\gamma=1$, and
frontier locality can degrade toward full-graph work.

The paper is organized as follows. Section~II defines the dynamic modularity
problem and notation. Section~III presents \method{} and its locality,
hierarchy, parallel-move, and complexity arguments. Section~IV reports the
experimental protocol and results. Section~V reviews modularity-based,
dynamic, and parallel community-detection methods, and Section~VI concludes.

\section{Problem and Preliminaries}

Let $\graph_t=(V_t,E_t,w_t)$ be the weighted directed graph observed after
batch $t$; undirected graphs are represented by a symmetric adjacency-weight
matrix, so $w_t(i,j)=w_t(j,i)$. A \emph{partition} $\partn_t$ is a set of
disjoint nonempty communities whose union is $V_t$. A batch update $\batch_t$
is a finite signed set of edge-weight increments,
\[
    w_t(i,j)=w_{t-1}(i,j)+\delta_{ij}, \quad (i,j,\delta_{ij})\in\batch_t,
\]
where positive and negative $\delta_{ij}$ represent insertions, deletions, or
weight changes, and zero-weight edges are removed. Let $S_t$ denote the
endpoints of changed edges and let $q_t=|S_t|$.

\textbf{Dynamic update problem.}
Given $(\graph_{t-1},\partn_{t-1},\batch_t)$ and the hierarchy produced by
the previous Leiden optimization, compute a partition $\partn_t$ of
$\graph_t$ that remains close to a Leiden-quality modularity optimum while
avoiding a full reoptimization of all vertices and aggregate levels after
each batch. The central algorithmic requirement is that the work should be
controlled by the part of the maintained state whose modularity context can
change, not by $|V_t|$ or $|E_t|$ whenever the update is local. The worst case
remains global: updates that change many communities require the algorithm to
visit the corresponding region of the hierarchy.

\textbf{Local update difficulty.}
Dynamic graph algorithms usually exploit the fact that an edge update changes
only a small part of the input graph~\cite{Rossetti2018,Hanauer2022}. For
community detection based on modularity, this locality is not purely
topological. Modularity uses a global null model and community-level degree
totals~\cite{NewmanGirvan2004,Brandes2008}; consequently, a local edge update
can alter the gain of moving vertices that belong to an affected community
even if their own incident edges did not change. The resolution-limit
literature gives a related structural view: modularity decisions are coupled
through aggregate edge volume and community scale
~\cite{Fortunato2010,Traag2011}. A dynamic Leiden update must therefore
identify not only changed ground vertices, but also the communities and
aggregate weights whose cached statistics participate in subsequent gain
computations.

For directed graphs, \method{} optimizes weighted modularity
\begin{equation}
\quality(\partn)=
\frac{1}{m}\sum_{i,j}
\left(w_{ij}-\gamma\frac{k_i^{out}k_j^{in}}{m}\right)
\mathbf{1}[c_i=c_j],
\label{eq:modularity}
\end{equation}
where $m=\sum_{ij}w_{ij}$, $k_i^{out}=\sum_jw_{ij}$,
$k_j^{in}=\sum_iw_{ij}$, $c_i$ is the community label of $i$, and
$\gamma=1$ in the experiments. The undirected objective is the symmetric
specialization of~\eqref{eq:modularity}. For a community $C$, let
$K_C^{out}=\sum_{i\in C}k_i^{out}$, $K_C^{in}=\sum_{i\in C}k_i^{in}$, and
$e_C=\sum_{i,j\in C} w_{ij}$ be its internal edge weight. These statistics
determine all one-step modularity gains considered by Leiden and are cached
by \method{} at each hierarchy level. Table~\ref{tab:notation} summarizes
the notation used below.

\begin{table}[!h]
\centering
\caption{Notation.}
\label{tab:notation}
\scriptsize
\begin{tabular}{@{}p{0.30\columnwidth}p{0.64\columnwidth}@{}}
\toprule
Symbol & Meaning \\
\midrule
$\graph_t^r=(V_t^r,E_t^r,W_t^r)$ & level-$r$ graph; $\graph_t^0=\graph_t$ \\
$R_r(u)$ & ground vertices represented by level-$r$ node $u$ \\
$\hier_t$ & hierarchy $(\graph_t^0,\ldots,\graph_t^L)$ and parent maps \\
$\pi_r(u)$ & parent supernode of level-$r$ node $u$ \\
$\partn_t$, $P_t^r$ & ground and level-$r$ partitions \\
$S_t$, $\frontier_t^r$ & changed endpoints and active level-$r$ frontier \\
$E_t^r(\frontier_t^r)$ & level-$r$ edges incident to active nodes \\
$q_t$, $d_r$ & $|S_t|$ and maximum degree at level $r$ \\
$M$, $L$, $N$ & move/refinement, hierarchy, and outer iteration budgets \\
$\alpha$, $\beta$ & gain and moved-frontier stopping thresholds \\
$p$, $s_t$ & thread count and per-thread temporary storage bound \\
\bottomrule
\end{tabular}
\end{table}

\section{\method}

\subsection{Incremental Hierarchical State}

\method{} maintains an inner hierarchy
$\hier_t=(\graph_t^0,\graph_t^1,\ldots,\graph_t^L)$. Level $0$ contains the
ground graph. A node at level $r+1$ is a supernode representing a refined
community of level $r$. Edges are stored only within the same level, and
parallel edges induced by aggregation are summed. The parent map
$\pi_r:V_t^r\rightarrow V_t^{r+1}$ records which supernode contains each
level-$r$ node; community labels at level $r$ are stored separately as
$P_t^r$. The separation records the distinction between the communities used
for local movement and the refined communities used to construct the next
aggregate level.

The hierarchy is the main structural difference from a simple warm-started
Leiden run. When $\batch_t$ changes level-0 weights, \method{} updates degree
totals, marks $S_t$, and follows parent links to the communities whose
aggregate incident weights may have changed. Unaffected supernodes retain
their parent, community, degree, and internal-weight statistics. Consequently,
the maintained hierarchy is both the search state used by Leiden-style
aggregation and the index that keeps future batches local.

The maintained hierarchy has two complementary roles. First, it preserves the
coarse representation on which Leiden obtains its quality: each supernode is
not only a cached label but also a contracted community whose incident aggregate
weights are available for subsequent local moves. Second, it is a routing
structure for dynamic repair. A changed ground edge can affect only the
communities containing its endpoints, their refined descendants, and the
aggregate edges induced by their parent pairs until later accepted moves
enlarge the frontier. The representation permits reuse of a stable high-level
partition while retaining the ability to expand beyond the immediate
ground-level neighborhood. The experiments therefore count hierarchy
maintenance in every reported runtime rather than treating the aggregate graph
as an offline preprocessing artifact.

\subsection{Hierarchical Locality}

The hierarchy changes the meaning of locality. Classical dynamic locality is
often described on the ground graph: an update is processed inside a
bounded-radius neighborhood of $S_t$. Such a definition is insufficient for a
Leiden-style objective: a level-$r$ supernode may represent many ground
vertices whose modularity context changes after a small number of aggregate
edge repairs. \method{} therefore defines locality at every hierarchy level.
At level $r$, a node $u\in V_t^r$ represents the ground set
$R_r(u)\subseteq V_t$ induced by the parent chain. Level $0$ nodes represent
singletons; higher-level nodes represent refined communities and their
aggregates.

\begin{figure*}[t]
    \centering
    \includegraphics[width=\textwidth]{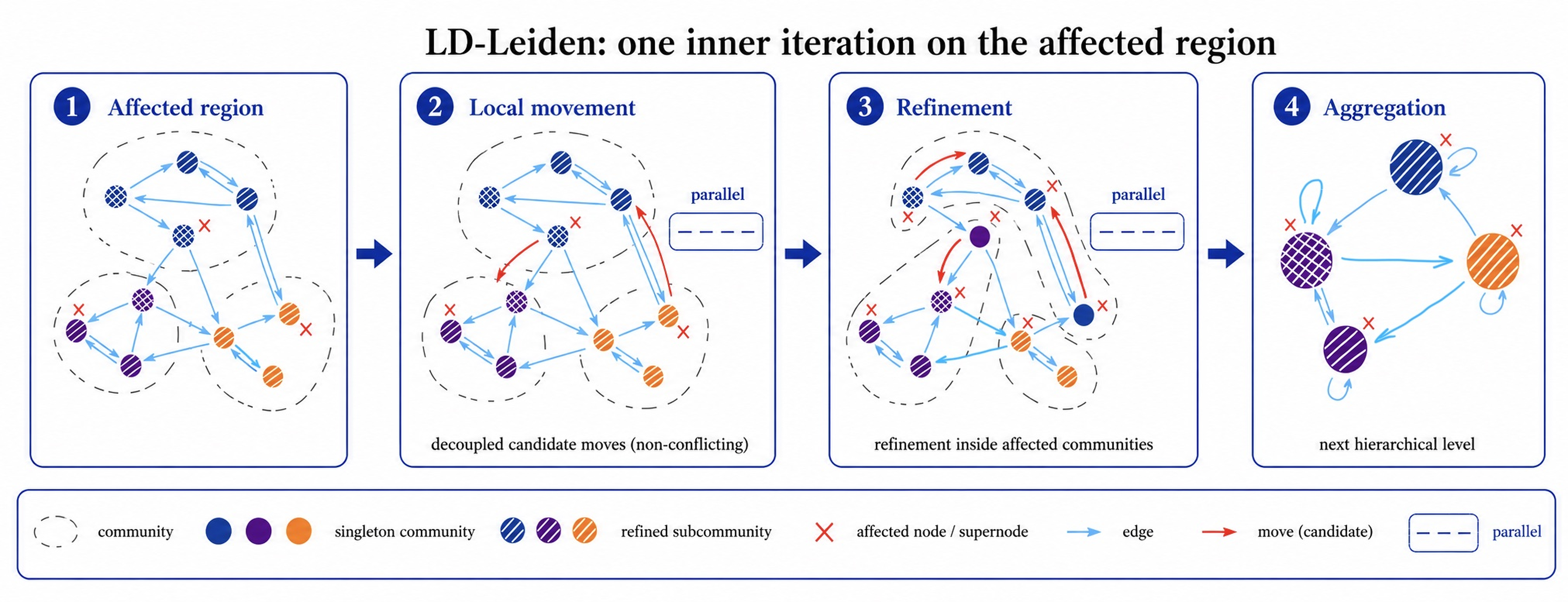}
    \caption{\method{} inner iteration on an affected region. The batch first
    identifies changed nodes and their incident communities, then performs
    frontier-restricted local movement with decoupled candidate moves,
    refinement inside affected communities, and aggregation to the next
    hierarchy level.}
    \label{fig:pipeline}
\end{figure*}

A node $u\in V_t^r$ is \emph{active} if it represents a changed endpoint,
belongs to a community whose cached incident aggregate weights changed, or is
reached by a move or refinement expansion from such a node. The active set is
$\frontier_t^r$, and $E_t^r(\frontier_t^r)$ denotes the level-$r$ edges
incident to active nodes. \method{} is local in the hierarchy in the sense
that movement, refinement, and aggregate repair are applied to $\frontier_t^r$ and
$E_t^r(\frontier_t^r)$ at each visited level, and the next frontier is
propagated only through parent links and changed aggregate edges. The
corresponding ground coverage may be large,
\[
    R(\frontier_t^r)=\bigcup_{u\in\frontier_t^r} R_r(u),
\]
but the processed hierarchy state can remain small. The resulting benefit is
that the algorithm can manipulate large ground-level groups through
supernodes, as required by the nonlinear modularity objective, while avoiding
aggregate communities whose incident weights and one-step modularity gains are
unchanged. In the worst case
$\frontier_t^r=V_t^r$; in sparse streams the experiments show much smaller
visited frontiers. Figure~\ref{fig:pipeline} illustrates the resulting
frontier-restricted Leiden iteration.

\subsection{Batch Update Procedure}

Each processed hierarchy level executes the Leiden stages, but only on the
current frontier. Local movement scans frontier vertices and considers moves
to adjacent communities. Refinement revisits only communities touched by
accepted moves and permits moves inside the current parent community, including
singleton refined communities. Aggregation then repairs the parent supernodes
and propagates the next frontier upward.

\begin{table}[t]
\centering
\refstepcounter{algorithm}
\label{alg:batch-update}
{\footnotesize ALGORITHM~\thealgorithm\\[-1pt]\textsc{\method{} batch update.}}\par
\vspace{2pt}
\scriptsize
\setlength{\tabcolsep}{1pt}
\renewcommand{\arraystretch}{0.92}
\begin{tabular*}{\columnwidth}{@{}c@{\;}c@{\;}r@{\;}l@{\extracolsep{\fill}}@{}}
\toprule
\multicolumn{4}{@{}l}{\textbf{Input:} $\hier_{t-1},\partn_{t-1},\batch_t,N,L,M,\alpha,\beta$} \\
\multicolumn{4}{@{}l}{\textbf{Output:} $\hier_t,\partn_t$} \\
\midrule
\algblock{3}{Update} & \algbrace{3}{6.3ex} & 1 & Apply $\batch_t$ to $\graph_{t-1}^0$; set changed endpoints $S_t$. \\
& & 2 & Repair endpoint degrees and dirty caches. \\
& & 3 & $\frontier_t^0\gets\textsc{InitFrontier}(S_t,\hier_{t-1},\batch_t)$. \\
\addlinespace[2.5pt]
\algblock{3}{Levels} & \algbrace{3}{6.3ex} & 4 & \textbf{for} $i=1,\ldots,N$ \textbf{do} over outer passes. \\
& & 5 & \hspace{.8em}\textbf{for} $r=0,\ldots,L$, $\frontier_t^r\ne\emptyset$ \textbf{do}. \\
& & 6 & \hspace{1.6em}$U\gets\frontier_t^r$; $Q_0\gets\quality(P_t^r)$. \\
\addlinespace[2.5pt]
\algblock{9}{Move} & \algbrace{9}{18.9ex} & 7 & \hspace{1.6em}\textbf{for} $j=1,\ldots,M$ \textbf{do} local movement. \\
& & 8 & \hspace{2.4em}$S\gets U\cup N_r(U)$ using level-$r$ neighbors. \\
& & 9 & \hspace{2.4em}$B\gets\textsc{Decouple}(\textsc{BestMove}_r(S))$. \\
& & 10 & \hspace{2.4em}Commit $B$; repair community totals. \\
& & 11 & \hspace{2.4em}\textbf{if} $\Delta\quality(B)<0$ \textbf{then} rollback; \textbf{break}. \\
& & 12 & \hspace{2.4em}$V\gets\textsc{Moved}(B)$; $U\gets\textsc{Expand}_r(V)$. \\
& & 13 & \hspace{2.4em}\textbf{if} $\Delta\quality(B)<\alpha Q_0$ \textbf{then break}. \\
& & 14 & \hspace{2.4em}\textbf{if} $|V|<\beta |S|$ \textbf{then break}. \\
& & 15 & \hspace{1.6em}\textbf{end for}. \\
\addlinespace[2.5pt]
\algblock{7}{Refine} & \algbrace{7}{14.8ex} & 16 & \hspace{1.6em}$R\gets\textsc{TouchedRefined}(U)$. \\
& & 17 & \hspace{1.6em}\textbf{for} $j=1,\ldots,M$ \textbf{do} refinement. \\
& & 18 & \hspace{2.4em}$B\gets\textsc{Decouple}(\textsc{BestRef}_r(R))$. \\
& & 19 & \hspace{2.4em}Commit $B$; $V\gets\textsc{Moved}(B)$. \\
& & 20 & \hspace{2.4em}\textbf{if} $|V|<\beta |R|$ \textbf{then break}. \\
& & 21 & \hspace{2.4em}$R\gets\textsc{SPExpand}_r(V)$ within parent community. \\
& & 22 & \hspace{1.6em}\textbf{end for}. \\
\addlinespace[2.5pt]
\algblock{2}{Agg.} & \algbrace{2}{4.4ex} & 23 & \hspace{1.6em}Repair $\pi_r$ and changed aggregate weights. \\
& & 24 & \hspace{1.6em}$\frontier_t^{r+1}\gets\textsc{NextFrontier}_r(U,R,\pi_r,\Delta W_t^r)$. \\
\addlinespace[2.5pt]
\algblock{4}{Stop} & \algbrace{4}{8.6ex} & 25 & \hspace{.8em}\textbf{end for}. \\
& & 26 & \hspace{.8em}\textbf{if} all frontiers are empty \textbf{then break}. \\
& & 27 & \hspace{.8em}\textbf{if} outer gain is below threshold \textbf{then break}. \\
& & 28 & \textbf{end for}; return level-0 labels induced by $\hier_t$. \\
\bottomrule
\end{tabular*}
\end{table}

Algorithm~\ref{alg:batch-update} summarizes the update procedure analyzed below.
Parallel stages read a fixed snapshot of community totals and emit tentative
changes into per-thread buffers; a compact serial phase filters conflicts
and commits accepted mini-batches. \textsc{BestMove} searches adjacent
communities of active nodes using Eq.~\eqref{eq:gain}. \textsc{BestRef}
restricts candidates to the current parent community and singleton refined
targets. \textsc{InitFrontier} builds the level-0 frontier from changed
endpoints, incident communities, and caches dirtied by $\batch_t$.
\textsc{Expand} adds moved nodes, their level-$r$ neighbors, touched
communities, and touched parent supernodes; \textsc{SPExpand} is the analogous
same-parent expansion used during refinement. \textsc{NextFrontier} maps active
or refined nodes and changed aggregate-weight endpoints through $\pi_r$ to seed
level $r+1$.
After every commit, cached
totals and affected aggregate weights are repaired before the next candidate
search. The procedure does not invoke a global Leiden pass; information
propagates through the frontier returned by movement, refinement, and
aggregate repair.

\subsection{Local Gain Reuse}

For a candidate move of vertex $v$ from community $a$ to community $b$, define
$w(v,C)=\sum_{u\in C}w_{vu}$ and $w(C,v)=\sum_{u\in C}w_{uv}$. Up to terms
independent of the target, the directed modularity gain is
\begin{align}
\Delta\quality(v:a\!\rightarrow\!b)
&= \frac{\Delta w_{vab}}{m}
 - \gamma\frac{\Delta d_{vab}}{m^2},
\label{eq:gain}\\
\Delta w_{vab}
&=w(v,b)+w(b,v)-w(v,a)-w(a,v), \nonumber\\
\Delta d_{vab}
&=k_v^{out}(K_b^{in}-K_a^{in})
 + k_v^{in}(K_b^{out}-K_a^{out}). \nonumber
\end{align}
Thus one-step gain evaluation in the current graph needs only incident weights
of $v$, its source and target community totals, and the current global total
$m$. After applying $\batch_t$, \method{} first repairs $m$, endpoint degrees,
dirty community totals, and affected aggregate weights, and then evaluates
candidate moves using Eq.~\eqref{eq:gain} on the repaired state. The frontier
screen is therefore a frontier-selection rule rather than an invariance claim:
vertices are rescanned when their incident weights, neighboring community
totals, or hierarchy context can change under the current batch. When a move is
accepted, \method{} adds graph neighbors, the old and new communities, and the
corresponding supernodes to the next frontier, since the accepted move may
change their gain contexts.

\subsection{Decoupled Parallel Moves}

Candidate discovery is parallel over frontier vertices. Each candidate is a
tuple $(v,c_{old},c_{new},\Delta\quality_v)$ with positive gain under the
current cached state. Candidates are sorted by decreasing gain and greedily
filtered. A candidate is accepted only if it does not leave a community already
marked as an acceptor and does not enter a community already marked as an
emitter. Hence each participating community has a single role in one apply
step: emitter, acceptor, or untouched.

For an accepted mini-batch, no community is both a source and a target. The
membership and total-weight updates of accepted moves can therefore be
accumulated without opposite-direction community conflicts, after which all
gain contexts are recomputed before the next mini-batch. The filter marks
every accepted source community as an emitter and every accepted target
community as an acceptor. It rejects any later candidate that would enter an
emitter or leave an acceptor, so no accepted community has both roles. Updates
inside a single role are additive; after they are applied, \method{} refreshes
the cached statistics before computing new candidates.

The filter is intentionally conservative: hub communities can attract many
candidate moves, causing rejections and repeated search after cache repair. The
scaling experiments include this cost, but do not isolate rejection rates.

\subsection{Incremental Aggregate Repair}

For level $r$, the aggregate edge weight between supernodes $x$ and $y$ at
level $r+1$ is
\begin{equation}
    W_{r+1}(x,y)=
    \sum_{\pi_r(u)=x,\ \pi_r(v)=y} W_r(u,v).
    \label{eq:aggregate}
\end{equation}
When only a set of level-$r$ nodes changes parent or incident weight,
\method{} subtracts their old incident contributions and adds the new ones.
After this subtract-add repair on all changed level-$r$ nodes and their
incident edges, every aggregate edge incident to an affected supernode has the
same weight as it would after recomputing level $r+1$ from level $r$.
Aggregate edges not incident to affected supernodes are unchanged.

This follows directly from Eq.~\eqref{eq:aggregate}, which is a grouped sum
over level-$r$ edges. For an unchanged node with unchanged parent and incident
weights, every summand remains in the same aggregate edge with the same value.
For a changed node, removing its old incident summands and inserting its new
summands exactly replaces the terms whose parent pair changed. Linearity gives
the same affected aggregate weights as full recomputation.

\subsection{Complexity and Memory}

Let $d_r$ be the maximum degree at level $r$. During one move/refinement
iteration, the frontier can expand through graph neighbors and through the
communities whose totals changed. With a fixed move/refinement budget $M$, the
number of level-$r$ nodes visited from an initial affected set of size $q_t$ is
bounded by $n_r\le d_r^{2M}q_t$ in the worst case. Candidate discovery scans
incident edges of visited nodes, and decoupling sorts at most one best
candidate per visited node. The resulting analysis gives the following
instance-dependent bound.

\begin{theorem}[Frontier-bounded update cost]
\label{thm:cost}
For fixed outer budget $N$, hierarchy budget $L$, and move/refinement budget
$M$, one \method{} batch update takes
\[
O\!\left(
N\sum_{r=0}^{L}
\left(|E_t^r(\frontier_t^r)|+|\frontier_t^r|\log |\frontier_t^r|\right)
\right)
\]
time. In the bounded-degree worst case at level $r$,
\[
|\frontier_t^r|\le n_r\le d_r^{2M}q_t,
\]
so a level iteration is $O(n_r(d_r+\log n_r))$.
\end{theorem}
\begin{proof}
At a processed level, move and refinement scans examine edges incident to
frontier nodes, which gives $O(|E_t^r(\frontier_t^r)|)$ work under fixed
iteration budgets. The decoupling filter sorts at most one retained candidate
per visited node, giving $O(|\frontier_t^r|\log|\frontier_t^r|)$. Applying
accepted moves and repairing aggregate edges revisits changed incident edges
only a constant number of times. Summing over levels and outer iterations
gives the first bound. The second follows by expanding from $q_t$ affected
endpoints through at most $2M$ neighbor/community steps, each with branching
factor at most $d_r$, and then substituting
$|E_t^r(\frontier_t^r)|\le n_rd_r$.
\end{proof}

The second form of the bound is conservative. In scale-free graphs, $d_r$ may
be comparable to $|V_t^r|$, so $d_r^{2M}q_t$ can degenerate to a full-level
bound. The operative claim is the first, frontier-dependent bound; the
experiments therefore report batch-size and average-degree behavior rather than
treating the maximum-degree worst case as tight.

The hierarchy stores per-node community, parent, degree, and internal-weight
arrays, plus adjacency lists at each maintained level. Let $n$ and $m$ be the
ground node and edge counts, $l$ the number of stored levels, and $p$ the
number of threads. In the implemented layout, the same accounting used to size
the containers gives the conservative byte bound
\[
\mathrm{Mem}_{total}\le
(104n+12m)l+128n+485p+664.
\]
The constants reflect fixed-width node records, edge records, community
records, and small global/thread-local containers. Asymptotically, this is
$O((n+m)l+n+p)$. The concrete bound makes the memory comparison in
Table~\ref{tab:aggregate} refer to the implemented hierarchy representation
rather than only to asymptotic storage.

\subsection{Relationship to Full Leiden}

\method{} is a dynamic optimizer, not a proof that local search reconstructs
the same partition as a full Leiden rerun. Greedy modularity optimization is
path-dependent, and a full rerun may choose a distant move sequence. The
guarantee used here is narrower: gain computations use repaired current-state
statistics, potentially changed contexts are brought into the frontier, and
affected aggregate weights are repaired exactly. The experiments then test
whether this local Leiden trajectory remains close to the full-quality
reference while reducing the work of repeated full-snapshot optimization.

\section{Experiments}
\label{sec:experiments}

\subsection{Setup}

The evaluation includes Leidenalg, DF-Leiden, NetworKit, and Grappolo.
Leidenalg is the warm-started static Leiden reference: after each batch, the
previous partition initializes the next optimization. DF-Leiden is the closest
dynamic Leiden baseline available under the matched protocol. NetworKit and
Grappolo are strong shared-memory Louvain baselines. Table~\ref{tab:baselines}
summarizes the empirical comparison surface. Directed weighted streams are
compared directly against Leidenalg, while the all-baseline comparison is
restricted to the common undirected subset, since the measured multicore
baselines do not expose the same directed interface. All measured methods use
the same batch files, graph representation interface, and matched iteration
budgets where applicable. Memory ratios are interpreted as implementation-level
measurements because node-id and edge-weight storage differ across systems.
The reported \method{} runtime includes maintenance of the internal hierarchy.

\begin{table}[t]
\centering
\caption{Baselines.}
\label{tab:baselines}
\scriptsize
\begin{tabular}{@{}lccccc@{}}
\toprule
Method & Core & Dynamic & Parallel & Directed & Weighted \\
\midrule
Leidenalg & Leiden & warm & no & yes & yes \\
DF-Leiden & Leiden & yes & yes & no & yes \\
NetworKit & Louvain & no & yes & no & yes \\
Grappolo & Louvain & no & yes & no & yes \\
\method{} & Leiden & yes & yes & yes & yes \\
\bottomrule
\end{tabular}
\end{table}

Experiments were run on a two-socket 96-core server with 512 GiB RAM, compiled
with GCC 10.2.1 and C++17. Unless otherwise stated, real-graph comparison
results use one thread and 100 batches. Temporal graphs in
Table~\ref{tab:datasets} are sorted by time and shuffled inside each batch;
static graphs are shuffled and streamed to stress large-scale update behavior.
Here $|E|$ is the number of unique edges and $|E'|$ is the temporal edge count
when timestamps are available. In Table~\ref{tab:datasets}, \emph{all}
marks undirected graphs used against all baselines, \emph{Leiden} marks
directed graphs used against the directed Leidenalg reference, and
\emph{scale} marks large undirected graphs used for thread and memory scaling.
Modularity ratios are final $\quality_{\method}/\quality_{baseline}$ after
all updates.

The protocol separates three experimental objectives that are often combined
in dynamic community-detection measurements. First, the Leidenalg comparison
tests whether a local method remains close to a high-quality Leiden reference
on directed and undirected streams. Second, the common undirected subset tests
performance against optimized shared-memory modularity systems under a matched
interface. Third, the batch-size, thread-scaling, memory, and synthetic
average-degree experiments test whether the dynamic mechanism scales with
update locality and available parallelism. All reported \method{} times
include graph updates, frontier construction, local movement, refinement,
decoupling, and aggregate hierarchy repair.

\begin{table}[t]
\centering
\caption{Datasets.}
\label{tab:datasets}
\scriptsize
\setlength{\tabcolsep}{2pt}
\begin{tabular}{@{}lrrcl@{}}
\toprule
Network & $|V|$ & $|E|/|E'|$ & Dir. & Eval. \\
\midrule
ca-cit & 28.1K & 3.15M/4.60M & no & all \\
wikipedia & 1.87M & 40.0M & yes & Leiden \\
flickr & 2.30M & 33.1M & yes & Leiden \\
stackoverflow & 2.60M & 36.2M/63.5M & yes & Leiden \\
youtube & 3.22M & 9.38M & no & all \\
com-lj & 4.00M & 34.7M & no & all+scale \\
com-orkut & 3.07M & 117M & no & all+scale \\
asia\_osm & 12.0M & 12.7M & no & all+scale \\
europe\_osm & 50.9M & 54.1M & no & all+scale \\
friendster & 68.4M & 2.59B & no & scale \\
dimacs10-uk & 105M & 3.30B & no & scale \\
kmer\_V1r & 214M & 465M & no & scale \\
\bottomrule
\end{tabular}
\vspace{4pt}
\end{table}

\subsection{Real-Graph Runtime and Quality}

The main experimental question is whether the incremental hierarchy reduces
work without degrading the modularity obtained by Leiden-style refinement.
Table~\ref{tab:aggregate} reports the aggregate evidence. Against
warm-started Leidenalg, \method{} is 48.77$\times$ faster across nine
100-batch streams and preserves a 0.996 mean final modularity ratio. It is
the most conservative quality reference in the study, as Leidenalg reruns the
full Leiden optimizer after every batch. On the six undirected graphs common to all
multicore baselines, \method{} is 43.45$\times$ faster than Grappolo,
9.73$\times$ faster than NetworKit, and 6.94$\times$ faster than DF-Leiden.
The final modularity ratios against these multicore baselines are 1.041,
1.031, and 1.027, respectively; thus the speedup is not obtained by abandoning
Leiden-level refinement. These timings include aggregate repair and hierarchy
maintenance, so the comparison is between complete dynamic updates rather than
isolated local-move kernels.

\begin{table}[t]
\centering
\caption{Aggregate results.}
\label{tab:aggregate}
\small
\begin{tabular}{@{}lrrrr@{}}
\toprule
Baseline & Graphs & Speedup & Mod. ratio & Mem. ratio \\
\midrule
Leidenalg & 9 & 48.77$\times$ & 0.996 & 0.597 \\
Grappolo & 6 & 43.45$\times$ & 1.041 & 0.643 \\
NetworKit & 6 & 9.73$\times$ & 1.031 & 0.509 \\
DF-Leiden & 6 & 6.94$\times$ & 1.027 & 0.586 \\
\bottomrule
\end{tabular}
\end{table}

\begin{figure}[t]
    \centering
    \includegraphics[width=\columnwidth]{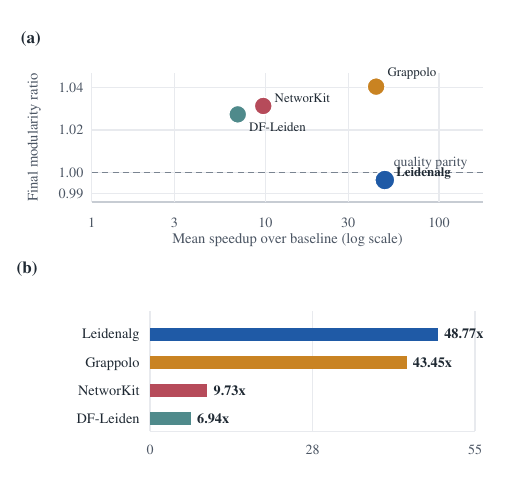}
    \caption{Runtime-quality summary on 100-batch real-graph updates. (a)
    Mean speedup against final modularity ratio
    $\quality_{\method}/\quality_{baseline}$; the dashed line marks quality
    parity. (b) The same mean speedups as bars.}
    \label{fig:runtime_quality}
\end{figure}

Figure~\ref{fig:runtime_quality} places these results on a speed-quality
plane. Leidenalg is the strictest reference in this plot, since it runs the
full Leiden optimizer after every batch. \method{} gives up less than 0.4\% mean final
modularity against this reference while reducing cumulative time by nearly two
orders of magnitude. Against DF-Leiden, NetworKit, and Grappolo, the point is
above the quality-parity line and to the right of runtime parity. The bar
panel reports the same comparison from the runtime side: the full-quality
reference is the slowest method, and the dynamic and multicore Louvain
baselines are also slower on the common graphs.

The per-dataset view separates sparse low-degree regimes from dense social
graphs. The largest Leidenalg gains occur on road and spatial graphs, where
the affected frontier tends to remain narrow after contraction. Social-network
gains are smaller as high-degree communities can pull more neighbors into the
frontier, but they remain substantial. The pattern matches
Theorem~\ref{thm:cost}: \method{} does not provide a graph-size independent
update, but replaces the full graph by the visited frontier and its aggregate
images. Preserved modularity across both families indicates that the hierarchy
repairs the context needed by Leiden-style refinement rather than performing an
overly narrow label update.

Table~\ref{tab:perdataset} gives the corresponding per-dataset evidence; each
cell reports speedup followed by the final modularity ratio
$\quality_{\method}/\quality_{baseline}$. Road and spatial
networks produce the largest gains, with 160.91$\times$ and 165.30$\times$
speedups against Leidenalg; low-degree structure keeps the hierarchy frontier
narrow even when the graph is very large. Dense social graphs expand the
frontier more often through high-degree communities, while \method{} still
avoids most global work. DF-Leiden is competitive on the smallest graph;
\method{} becomes substantially faster on larger sparse instances while also
improving final modularity. Directed datasets have dashes in the DF-Leiden
columns since the common all-baseline protocol is undirected.

\begin{center}
\begin{minipage}{\columnwidth}
\centering
\refstepcounter{table}
\label{tab:perdataset}
{\footnotesize TABLE~\thetable\\[-1pt]\textsc{Per-dataset results.}}\par
\vspace{2pt}
\scriptsize
\setlength{\tabcolsep}{1.6pt}
\begin{tabular}{@{}lrrr@{}}
\toprule
Network & Time & vs Leidenalg & vs DF-Leiden \\
\midrule
ca-cit-HepPh & 26.5 & 9.19$\times$/0.989 & 1.44$\times$/1.022 \\
wikipedia & 638.5 & 12.95$\times$/0.994 & -- \\
flickr & 406.7 & 17.47$\times$/0.996 & -- \\
stackoverflow & 592.7 & 23.76$\times$/0.995 & -- \\
youtube & 263.8 & 24.91$\times$/0.988 & 3.72$\times$/1.042 \\
asia\_osm & 167.2 & 160.91$\times$/1.004 & 16.57$\times$/1.035 \\
europe\_osm & 767.2 & 165.30$\times$/1.004 & 15.58$\times$/1.035 \\
com-lj & 856.6 & 13.87$\times$/1.000 & 2.40$\times$/1.014 \\
com-orkut & 1928.1 & 10.54$\times$/0.999 & 1.92$\times$/1.017 \\
\bottomrule
\end{tabular}
\end{minipage}
\end{center}

\begin{figure*}[t]
    \centering
    \includegraphics[width=\textwidth]{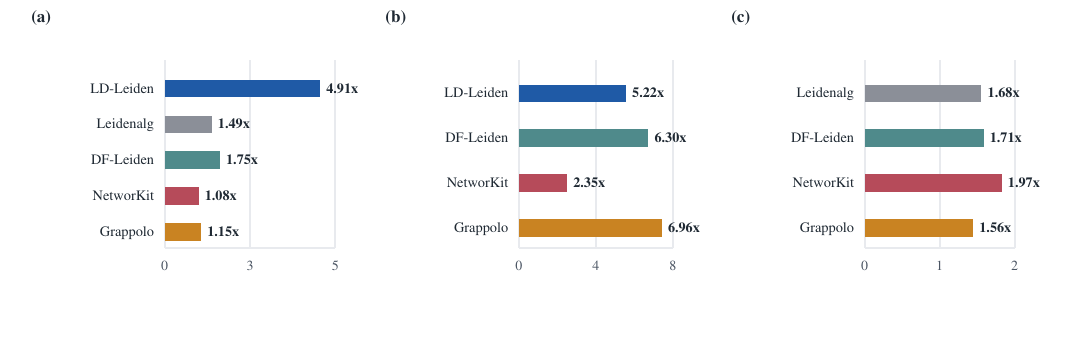}
    \caption{Batch-size sensitivity, thread scaling, and memory. (a) Mean
    runtime improvement from a 10$\times$ smaller batch. (b) Mean 64-thread
    speedup over one thread. (c) Baseline resident memory divided by
    \method{} memory on common graphs.}
    \label{fig:scalability_batch}
\end{figure*}

\subsection{Batch Size, Parallelism, and Memory}

\begin{table}[!b]
\centering
\caption{Batch regimes.}
\label{tab:batch_modes}
\vspace{2pt}
\scriptsize
\setlength{\tabcolsep}{1.9pt}
\resizebox{\columnwidth}{!}{%
\begin{tabular}{@{}lrrrr@{}}
\toprule
Baseline & 1 & 10 & 100 & 1000 \\
\midrule
Leidenalg & 2.12$\times$/1.007 & 7.50$\times$/0.997 &
48.77$\times$/0.996 & 365.54$\times$/0.988 \\
Grappolo & 4.58$\times$/1.031 & 10.25$\times$/1.038 &
43.45$\times$/1.041 & 73.65$\times$/1.016 \\
NetworKit & 0.22$\times$/1.022 & 1.35$\times$/1.029 &
9.73$\times$/1.031 & 77.59$\times$/1.027 \\
DF-Leiden & 0.50$\times$/1.067 & 1.37$\times$/1.037 &
6.94$\times$/1.027 & 44.16$\times$/1.019 \\
\bottomrule
\end{tabular}
}
\end{table}

Batch-size sensitivity separates local dynamic behavior from faster static
optimization. \method{} improves by 4.91$\times$ on average whenever the
batch size is reduced by a factor of ten. The
corresponding factors are 1.49$\times$ for Leidenalg, 1.75$\times$ for
DF-Leiden, 1.08$\times$ for NetworKit, and 1.15$\times$ for Grappolo. This
pattern is the empirical counterpart of Theorem~\ref{thm:cost}: the repeated
work follows the changed frontier and its aggregate images rather than the
full graph.

Table~\ref{tab:batch_modes} gives the broader batch-regime view; each entry
reports speedup followed by the final modularity ratio. With a single
large batch, \method{} behaves like a static high-quality optimizer and is not
always faster than highly optimized Louvain implementations. As the same graph
is split into finer updates, its locality becomes visible: speedup against
Leidenalg grows from 2.12$\times$ at one batch to 365.54$\times$ at 1000
batches, while the final modularity ratio remains 0.988 even in the finest
tested regime. The same trend appears against DF-Leiden, where speedup grows
from 0.50$\times$ at one batch to 44.16$\times$ at 1000 batches.

Figure~\ref{fig:scalability_batch} ties the locality result to parallelism and
memory. On nine undirected scaling graphs, \method{} obtains a mean
5.22$\times$ speedup from 1 to 64 threads and 1.33$\times$ per thread
doubling. The scaling follows the algorithm structure: candidate discovery,
move application, refinement, and affected aggregation run in parallel, while
the decoupling filter is the compact serial step. DF-Leiden and Grappolo have
comparable relative scaling, but they start from slower one-thread runs and
use more memory on the common graphs. NetworKit scales more weakly and has
memory limitations on the largest instances. In Table~\ref{tab:aggregate},
Mem. ratio is $\mathrm{RSS}_{\method}/\mathrm{RSS}_{baseline}$, so values below
one mean that \method{} uses less memory on matched inputs. This supports the
concrete hierarchy storage bound used with Theorem~\ref{thm:cost}.
In absolute terms, one-thread \method{} memory ranges from 0.51 GiB to
23.67 GiB on the nine main graphs; scale-only runs use 90.40/201.68 GiB on
kmer\_V1r and 372.76/425.52 GiB on dimacs10-uk with 1/64 threads.

\subsection{Modularity Variability and Parameter Selection}

Parallel local search may change the order in which admissible moves are
accepted. We therefore evaluate whether the final modularity reported in the
main tables is sensitive to the thread count used by the implementation. On the
scaling graphs where 1--64 thread runs are available, the relative range of
\method{} final modularity is 0.38\% on average and 1.49\% in the worst case.
The same check gives 5.11\% and 11.87\% for DF-Leiden, and 2.76\% and 5.64\%
for NetworKit, respectively. Thus, for \method{}, changing the amount of
parallelism has only a small effect on the final objective value, while the
runtime gains in Section~\ref{sec:experiments} are not obtained by selecting an
unstable run.

A separate 10-run repeatability check on six common graphs separates this
thread-count effect from run-to-run variability. In the static protocol,
\method{} has a maximum reported interval of 0.00\%, whereas Leidenalg reaches
5.10\%. In the dynamic protocol, the maximum interval for \method{} is 0.68\%;
the corresponding maxima are 0.23\% for Leidenalg, 0.80\% for NetworKit, and
1.50\% for DF-Leiden. These values indicate that the dynamic quality reported
for \method{} is stable under repeated execution as well as under changes in
parallelism. Figure~\ref{fig:robustness-tuning} summarizes these stability
checks together with the parameter-selection sweep.

\begin{center}
\begin{minipage}{\columnwidth}
\centering
\refstepcounter{figure}
\includegraphics[width=\columnwidth]{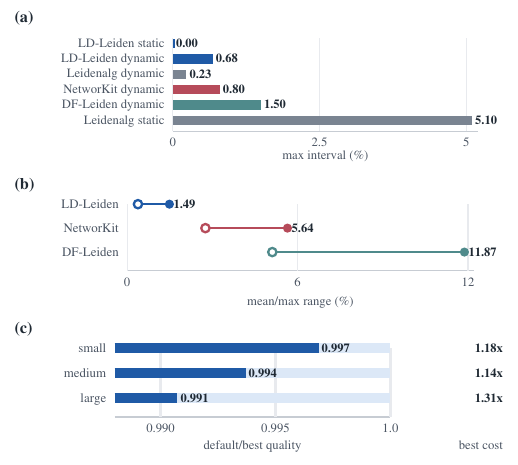}
\inlinefigcaption{Robustness and parameter-selection checks. (a) Maximum
modularity intervals over repeated runs. (b) Final-modularity
sensitivity to thread count; open and filled markers denote mean and maximum
ranges. (c) Default quality relative to the best observed configuration, with
right-side labels giving the runtime of the best-quality configuration
relative to the default.}
\label{fig:robustness-tuning}
\end{minipage}
\end{center}

Parameter selection uses the same quality-constrained criterion on small,
medium, and large graph groups. Each sweep evaluates 7203 configurations over
$N,L,M,\alpha,\beta$ and yields 123, 70, and 111 non-dominated
Pareto-frontier points. The default setting, $N=2$, $L=4$, $M=5$,
$\alpha=0.001$, and $\beta=0.01$, retains 0.997, 0.994, and 0.991 of the best
observed modularity on the three groups. The corresponding best-modularity
configurations are 1.18$\times$, 1.14$\times$, and 1.31$\times$ slower than
the default. These sweeps support using one fixed configuration in all reported
experiments: the default is close to the quality frontier across graph scales
without per-dataset tuning.
All sweeps fix $\gamma=1$. Changing $\gamma$ changes community granularity and
the active frontier, so resolution sensitivity is not claimed here.

\begin{figure*}[!t]
    \centering
    \includegraphics[width=.96\textwidth]{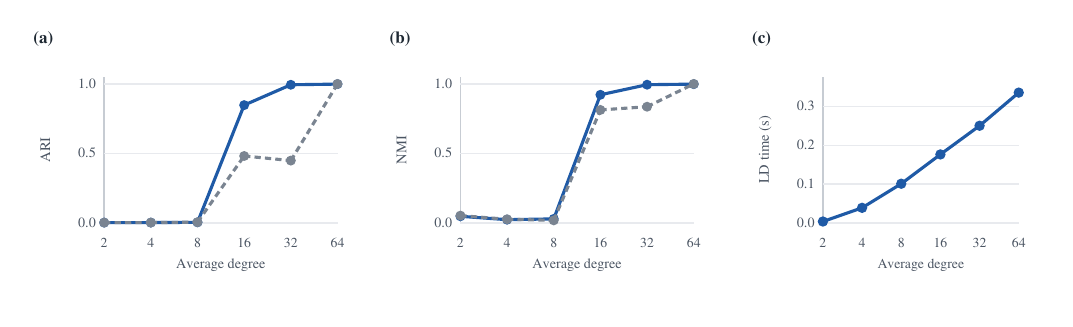}
    \caption{Synthetic SBM-style dynamic-sequence summary. In panels (a,b),
    solid blue denotes \method{} and dashed gray denotes Leidenalg. (a) ARI
    against planted labels. (b) NMI against planted labels. (c) \method{}
    dynamic-batch time versus average degree as a proxy for local edge volume.}
    \label{fig:synthetic}
\end{figure*}

\subsection{Synthetic Recovery}

The real-graph results measure modularity, but they do not reveal whether
locality changes planted-community recovery. The synthetic evaluation is based
on a dynamic SBM-style planted-partition benchmark. Each sequence starts from
a static SBM-style snapshot with planted labels, then applies controlled edge
rewiring and splits the resulting updates into dynamic batches of 500 edge
changes; batch 0 initializes the full snapshot. The sweep varies average
degree $\bar d$ from 2 to 64 and uses multiple maximum-degree and seed
settings.

When the graph is too sparse to identify the planted partition, both Leidenalg
and \method{} have near-zero ARI. Once
$\bar d$ reaches 16, \method{} recovers the planted communities more accurately
than warm-started Leidenalg under the tested dynamic schedule: mean ARI is
0.848 versus 0.481 at degree 16 and 0.994 versus 0.448 at degree 32. At degree
64, both methods recover the planted partition almost perfectly. Across these
settings, \method{} remains 9.0$\times$--250.6$\times$ faster per dynamic
batch.

The same synthetic sweep also checks the complexity mechanism.
Theorem~\ref{thm:cost} uses the maximum level degree $d_r$ as a worst-case
quantity. In the SBM-style generator, however, $\bar d$ is an
empirical proxy for the expected local edge volume scanned by the frontier.
The measured \method{} dynamic-batch time increases monotonically with
$\bar d$. On the identifiable-to-dense part of the sweep
($\bar d\ge8$), a log-log fit of time versus $\bar d$ has slope 0.57 and
$R^2=0.977$. The fit is empirical rather than an average-case theorem, but it
supports the intended interpretation of the local bound: runtime grows with
the local edge volume touched by the affected hierarchy, while the worst-case
max-degree expansion remains conservative.

Figure~\ref{fig:synthetic} separates structure recovery from update cost.
Panels (a) and (b) compare both algorithms against the planted labels by
ARI and NMI, showing that once the benchmark enters an identifiable regime,
\method{} preserves the recoverable community structure while using local
updates. Panel (c) then reports the cost of that complete dynamic-batch path,
including graph updates, frontier construction, movement, refinement,
decoupling, and aggregate repair, rather than only the local-move kernel.

\subsection{Interpretation and Limits}

The evidence identifies a clear operating envelope. \method{} is most attractive
when updates are small relative to the graph, the previous partition remains
meaningful, and Leiden-level modularity is required. Its advantage comes from
repairing the hierarchy touched by the affected frontier, including parent
ranges and aggregate edges, instead of rebuilding every level after every
batch. Maintaining this state increases cost relative to label-only
incremental heuristics, but it also enables comparison against Leiden-style
move-refine-aggregate quality rather than only against local label stability.

Locality is not a universal guarantee. A large batch or
an update that repeatedly touches high-degree bridge communities can make the
frontier approach a full update. Moreover, greedy modularity optimization is
path-dependent, so \method{} is not guaranteed to reproduce the exact
trajectory or partition of a full Leiden rerun; equal modularity values may
still correspond to different community assignments. For deployments where
community identity matters, periodic pairwise NMI/ARI checks against a
full-snapshot refresh are a practical diagnostic alongside frontier size,
aggregate repair volume, and realized gain. When locality indicators approach
the full graph, a full or partially full refresh becomes preferable.

\section{Related Work}

\textbf{Static modularity and Leiden.}
Community detection has structural, statistical, information-theoretic, and
dynamical formulations~\cite{GirvanNewman2002,Newman2003,Porter2009,
Fortunato2010,Rosvall2007,Mucha2010}. The present work belongs to the
modularity-optimization line, where the partition quality is defined by the
excess internal weight relative to a configuration-model null expectation
~\cite{NewmanGirvan2004,Brandes2008}. Modularity has well-known limitations,
including hard optimization problems and resolution-limit effects
~\cite{Brandes2008,Fortunato2010,Reichardt2006,Traag2011}; nevertheless, it
remains the objective optimized by Louvain, Leiden, related modularity
variants, and many scalable graph-mining benchmarks
~\cite{Clauset2004,Duch2005,Newman2006,Blondel2008,Lancichinetti2008,
Lancichinetti2009,Yang2016,Mucha2010,Traag2009,Traag2019}.
Louvain introduced the local-move and aggregation template, while Leiden adds
a refinement phase that guarantees well-connected communities on static
snapshots~\cite{Blondel2008,Traag2019}. \method{} preserves this
move-refine-aggregate structure. In contrast to label-propagation
variants~\cite{TraagSubelj2023} or label-only local updates, this setting
requires the dynamic state to maintain community volumes, refined
communities, and aggregate edge weights after each batch.

\textbf{Dynamic community detection.}
Dynamic surveys separate snapshot matching, temporal smoothing, local update,
stream clustering, and temporal probabilistic models~\cite{Rossetti2018}.
Early and representative methods reuse static solvers
~\cite{Aynaud2010}, process incremental batches~\cite{Chong2013}, track
communities online~\cite{Shang2014}, or update labels and modularity
statistics locally~\cite{Aktunc2015,Yin2016,Held2016,Meng2016}. DynaMo,
C-Blondel, and Delta-Screening are modularity-based examples in which the
previous partition restricts part of the subsequent search
~\cite{Zhuang2019,Seifikar2020,Zarayeneh2021}. A broader set of dynamic
methods addresses related but different objectives: label propagation
~\cite{Xie2013}, online community discovery~\cite{Rossetti2017Tiles},
local-community updates~\cite{Zakrzewska2015}, game-theoretic updates
~\cite{Jiang2015}, flow stability~\cite{Bovet2022}, temporal smoothness
and Matthew-effect dynamics~\cite{Sun2022,Cai2023}, social-network tracking
~\cite{Mazza2023}, sparse temporal data~\cite{Conrad2025}, random-walk
snapshot clustering~\cite{Blaskovic2025},
matrix-factorization-based temporal clustering~\cite{Yu2025}, spectral
reconstruction in time-evolving stochastic block models~\cite{DallAmico2020},
and Bayesian models for sparse overlapping dynamic communities
\cite{Laos2025}.
These formulations maintain different dynamic objects:
temporally stable labels, evolving local communities, flow-based partitions,
latent factors, or generative memberships. A Leiden-style dynamic modularity
problem additionally requires, after each batch, a partition of the current
weighted graph and the refined aggregate state from which later modularity
gains are evaluated.

\textbf{Dynamic Louvain and Leiden methods.}
Recent dynamic Louvain and Leiden methods reuse the previous partition, but
they differ in the state retained between batches and in whether they preserve
Leiden refinement. Frontier-based methods start from the previous partition,
expand from changed endpoints, and screen vertices whose local modularity
context is unlikely to change
~\cite{Sahu2024Louvain,Sahu2024Leiden,Sahu2024Tracking}. This line reduces
the repeated work of full snapshot optimization, but the maintained state is
mainly a partition and a local candidate set. Parallel multicore variants
increase throughput for dynamic modularity tracking, while still facing the
coupling among movement, refinement, and aggregation
~\cite{Sahu2025Parallel}.

A separate line treats the hierarchy itself as dynamic state. In this view,
aggregate graphs, parent links, refined communities, and changed supernodes
are maintained across batches instead of being reconstructed after every
snapshot. \method{} follows this hierarchical direction and focuses on local
aggregate repair, directed weighted updates, and decoupled shared-memory move
batches.

\textbf{Parallel modularity systems.}
Parallel modularity systems clarify the static performance context even when
they solve snapshots rather than dynamic updates. Grappolo and related work
accelerate Louvain-style modularity optimization~\cite{Halappanavar2017,
Lu2015}; NetworKit provides high-performance graph analytics and modularity
heuristics~\cite{Staudt2016}; distributed Louvain studies identify
synchronization and memory costs in large graph partitions
~\cite{Sattar2018,Ghosh2018}.
Shared-memory Leiden implementations and GVE-Leiden improve static Leiden
throughput~\cite{Ueckerdt2021,Verweij2020,Sahu2023GVE}; GPU-oriented gLeiden
is a complementary static Leiden system for directed and undirected
graphs~\cite{Gul2026GLeiden}. These systems optimize a complete snapshot,
whereas \method{} starts from the
previous hierarchy and repairs only the affected state. Scalability in this
setting depends on whether maintaining local, hierarchical state remains less
expensive than repeatedly optimizing high-quality snapshot partitions.

\textbf{Experimental context.}
Large-scale graph-mining studies commonly combine real network collections
with synthetic graphs that expose structure-recovery behavior. The real-graph
experiments in this paper use KONECT, SNAP, and Network Data Repository
graphs~\cite{Kunegis2013,Leskovec2016,Rossi2015}. The synthetic experiments
use SBM-style dynamic sequences related to standard planted-partition
models~\cite{Karrer2011,Decelle2011}. ARI and NMI quantify agreement with
known planted labels and complement modularity with an external
partition-similarity view~\cite{Hubert1985,Danon2005}.

\section{Conclusion}

This work addressed the problem of maintaining Leiden-quality modularity
partitions on large dynamic graphs without full-snapshot optimization after
each batch. Full reruns preserve the objective but scale with the whole graph;
existing dynamic alternatives often save work by weakening refinement,
retaining less hierarchy state, or narrowing graph support. \method{} addresses
this gap by retaining the Leiden move-refine-aggregate pipeline and making only
affected hierarchy state dynamic. Its innovation is the joint use of an
affected-frontier rule after statistic repair, exact subtract-add aggregate
repair, and conflict-filtered parallel local moves with additive community
updates. Thus, the analyzed work depends on the visited frontier and its
aggregate images rather than on the full snapshot, and the experiments show
large speedups while preserving or improving final modularity against strong
baselines. Worst-case global propagation remains a limitation and motivates
adaptive refresh rules that switch to full reoptimization when locality no
longer holds.

\end{document}